\documentclass{llncs}
\usepackage{amsmath,amssymb}
\usepackage{url}
\usepackage{floatflt}
\usepackage{paralist}
\usepackage{theorem}
\renewcommand{\subparagraph}{}
\usepackage{ifthen}
\usepackage[all]{xy}

\title{Probabilistic Anonymity and Admissible Schedulers}
\author{Flavio D. Garcia \and Peter van Rossum \and Ana Sokolova}
\institute{Institute for Computing and Information Sciences, \\
           Radboud University Nijmegen, The Netherlands. \\
           \texttt{http://www.cs.ru.nl/\~{ }\{flaviog, petervr, anas\}}}

\newcommand{\signed}%
    {{\unskip\nobreak\hfill\penalty50
      \hskip2em\hbox{}\nobreak\hfil $\square$
      \parfillskip=0pt \finalhyphendemerits=0 \par}}

    \newcounter{alphacnt}
    \newenvironment{alphalist}
    { \noindent
    \setcounter{alphacnt}{0}
    \begin{list}{(\alph{alphacnt})}
        {   \settowidth{\labelwidth}{(a)\ }
            \setlength{\leftmargin}{\labelwidth}
        \usecounter{alphacnt}}      }%
    { \end{list} }

\newcommand{\weg}[1]{}

\newcommand{\Act}{\mbox{\sl Act}}

\newcommand{\calA}{\mbox{$\mathcal A$}}

\newfont{\openface}{msbm10}

\mathchardef\ls="213C                   % less symbol (< used as \langle)
\mathchardef\gr="213E                   % greater symbol (> used as \rangle)
\newcommand{\after}{\mathbin{\lower -0.25ex \hbox{\mbox{\tiny $\circ$}}}}

\DeclareMathOperator{\supp}{supp} 

\DeclareMathOperator{\first}{first}
\DeclareMathOperator{\last}{last}

\DeclareMathOperator{\trace}{trace}

\DeclareMathOperator{\Paths}{Paths}

\newcommand{\compress}{false}

\ifthenelse{\boolean{\compress}}{
\usepackage[small,compact]{titlesec}
}{\usepackage[small]{titlesec}}

\newcommand{\Prob}{{\mathbb P}}

\newcommand{\vbar}{\;|\;}            % for use in grammars
\newcommand{\mnote}[1]{}
\renewcommand{\Act}{{\text{Act}}}
\newcommand{\ActO}{{\text{Act}_O}}

\newtheorem{thm}{Theorem}
\numberwithin{thm}{section}

\spnewtheorem{dfn}[thm]{Definition}{\bfseries}{\rmfamily}
\spnewtheorem{exa}[thm]{Example}{\bfseries}{\rmfamily}

\ifthenelse{\boolean{\compress}}{
\setlength{\theorempreskipamount}{4pt plus 1pt}
\setlength{\theorempostskipamount}{4pt plus 1pt}
\setlength{\parskip}{0pt plus 1pt}
}{}

\begin{document}

\setlength{\abovedisplayskip}{4pt plus 1pt minus 1pt}
\setlength{\belowdisplayskip}{4pt plus 1pt minus 1pt}

\maketitle
\begin{abstract}
When studying safety properties of (formal) protocol models, it is customary to view the scheduler as an adversary: an entity trying to
falsify the safety property. We show that in the context of security protocols, and in particular of anonymizing protocols, this gives the
adversary too much power; for instance, the contents of encrypted messages and internal computations by the parties should be considered
invisible to the adversary.

We restrict the class of schedulers to a class of admissible schedulers which better model adversarial behaviour. These admissible
schedulers base their decision solely on the past behaviour of the system that is visible to the adversary.

 Using this, we propose a definition of anonymity: for all admissible schedulers the identity of the users and the observations of the
adversary are independent stochastic variables. We also develop a proof technique for typical cases that can be used to proof anonymity: a
system is anonymous if it is possible to `exchange' the behaviour of two users without the adversary `noticing'.
\end{abstract}

\newcommand{\fpaths}{{\operatorname{{Paths}^{\leq\omega}}}}
\newcommand{\paths}{{\operatorname{{Paths}}}}
\newcommand{\cpaths}{{\operatorname{{CPaths}}}}
\newcommand{\sched}{{\operatorname{{Sched}}}}
\newcommand{\trdistr}{{\operatorname{\mathit{trdistr}}}}
\newcommand{\tree}{\operatorname{{tree}}}
\newcommand{\distr}{{\operatorname{{\mathcal{D}}}}}
\newcommand{\pso}{{\operatorname{\Omega}}}
\newcommand{\psf}{{\operatorname{\mathcal{F}}}}
\newcommand{\psp}{{\operatorname{\mathbf{P}}}}

\section{Introduction}

%This paper makes the following point: \\

%\begin{minipage}{11cm}{ \bf In the analysis of systems that include probabilities and nondeterminism in terms of probabilistic anonymity,
%not all schedulers (resolving nondeterminism) may be allowed}
%\end{minipage}\\

\noindent
Systems that include probabilities and nondeterminism are very convenient for modelling probabilistic (security) protocols. Nondeterminism is highly
desirable feature for modelling implementation freedom, action of the environment, or incomplete knowledge about the system.\\

\noindent It is often of use to analyze probabilistic properties of such systems as for example ``in 30\% of the cases sending a message is
followed by receiving a message'' or ``the system terminates successfully with probability at least 0.9''. Probabilistic
anonymity~\cite{bp_2005_probabilistic} is also such a property. In order to be able to consider such probabilistic properties we must first
eliminate the nondeterminism present in the models. This is usually done by entities called schedulers or adversaries. It is common in the
analysis of probabilistic systems to say that a model with nondeterminism and probability satisfies a probabilistic property
if and only if it satisfies it no matter in which way the nondeterminism was resolved, i.e., for \emph{all possible schedulers}.\\

\noindent On the other hand, in security protocols, adversaries or schedulers are malicious entities that try to break the security of the
protocol. Therefore, allowing just any scheduler is inadmissible. We show that the well-known Chaum's Dining Cryptographers (DC)
protocol~\cite{cha_1988_dining} is not anonymous if we allow for all possible schedulers. Since the protocol is well-known to be anonymous,
this shows that for the treatment of
probabilistic security properties, in particular probabilistic anonymity, the general approach to analyzing probabilistic systems does not directly fit.\\

\noindent We propose a solution based on restricting the class of all schedulers to a smaller class of \emph{admissible schedulers}. Then
we say that a probabilistic security property holds for a given model, if the property holds after resolving the nondeterminism under
\emph{all admissible schedulers}.

\section{Probabilistic Automata}

In this section we gather preliminary notions and results related to probabilistic automata~\cite{SL94:concur,Seg95:thesis}. Some of the
formulations we borrow from~\cite{Sok05:thesis} and~\cite{Che06:thesis}. We shall model protocols with probabilistic automata. We start
with a definition of probability distribution.

\begin{dfn}[Probability distribution]\label{PrDisDef}
A function $\mu \colon S \to [0,1]$ is a discrete probability distribution, or distribution for short, on a set~$S$ if $\sum_{x \in S}
\mu(x) = 1$. The set $\{x \in S|\ \mu(x) \gr 0\}$ is the support of~$\mu$ and is denoted by~$\supp(\mu)$. By $\mathcal{D}(S)$
 we denote the set of all discrete probability distributions on the set $S$.
\end{dfn}

\noindent We use the simple probabilistic automata~\cite{SL94:concur,Seg95:thesis}, or MDP's~\cite{Bellman_1957_markov} as models of our probabilistic
processes. These models are very similar to the labelled transition systems, with the only difference that the target of each transition is
a distribution over states instead of just a single next state.

\begin{dfn}[Probabilistic automaton]\label{ProbAutDef}
A probabilistic automaton is a triple $\calA = \langle S , A , \alpha \rangle$ where:
\begin{itemize}
\item $S$ is a set of states.
\item $A$ is a set of actions or action labels.
\item $\alpha$ is a transition function $\alpha: S \to {\mathcal{P}}(A \times \distr S)$.
\end{itemize}
A terminating state of $\mathcal A$ is a state with no outgoing transition, i.e. with $\alpha(s) = \emptyset$. We might sometimes also
specify an initial state $s_0 \in S$ of a probabilistic automaton $\calA$. We write $s \stackrel{a}{\to} \mu$ for $(a,\mu) \in \alpha(s), \
s\in S$. Moreover, we write $s \stackrel{a,\mu}\leadsto t$ for $s, t \in S$ whenever $s \stackrel{a}{\to} \mu$ and $\mu(t) \gr 0$.
\end{dfn}

\noindent We will also need the notion of a fully probabilistic system.

\begin{dfn}[Fully Probabilistic Automaton]\label{FProbAutDef}
A fully probabilistic automaton is a triple $\calA = \langle S , A , \alpha \rangle$ where:
\begin{itemize}
\item $S$ is a set of states.
\item $A$ is a set of actions or action labels.
\item $\alpha$ is a transition function $\alpha: S \to \distr(A \times S) + 1$.
\end{itemize}
Here $1 = \{*\}$ denotes termination, i.e., if $\alpha(s) = *$ then $s$ is a terminating state. It can also be understood as a
zero-distribution i.e. $\alpha(s)(a,t) = 0$ for all $a \in A$ and $t \in S$. By $s_0 \in S$ we sometimes denote an initial state of
$\calA$. We write $s {\to} \mu$ for $\mu = \alpha(s), \ s\in S$. Moreover, we write $s \stackrel{a}\leadsto t$ for $s, t \in S$ whenever $s
{\to} \mu$ and $\mu(a,t) \gr 0$.
\end{dfn}

\noindent A major difference between the (simple) probabilistic automata and the fully probabilistic ones is that the former can express nondeterminism.
In order to reason about probabilistic properties of a model with nondeterminism we first resolve the nondeterminism with help of schedulers or
adversaries -- this leaves us with a fully probabilistic model whose probabilistic behaviour we can analyze. We explain this in the sequel.\\

\begin{dfn}[Paths]\label{PathsDef}
A path of a \emph{probabilistic automaton} $\mathcal A$ is a sequence
$$\pi = s_0 \stackrel{a_1,\mu_1}{\to} s_1  \stackrel{a_2,\mu_2}{\to} s_2 \dots$$
where $s_i \in S$, $a_i \in A$
and $s_i \stackrel{a_{i+1},\mu_{i+1}}{\leadsto} s_{i+1}$.

A path of a \emph{fully probabilistic automaton} $\mathcal A$ is a sequence
$$\pi = s_0 \stackrel{a_1}{\to} s_1  \stackrel{a_2}{\to} s_2 \dots$$
where again $s_i \in S$, $a_i \in A$
and $s_i \stackrel{a_{i+1}}{\leadsto} s_{i+1}$.

A path can be finite in which case it ends with a state. A path is complete if it is either infinite or finite ending in a terminating
state. We let $\last(\pi)$ denote the last state of a finite path $\pi$, and for arbitrary path $\first(\pi)$ denotes its first state. A
trace of a path is the sequence of actions  in $A^{*} \cup A^{\infty}$ obtained by removing the states (and the distributions), hence above
$\trace(\pi) = a_1a_2\ldots$. The length of a finite path $\pi$, denoted by $|\pi|$ is the number of actions in its trace. Let
$\paths(\mathcal A)$ denote the set of all paths, $\fpaths(\mathcal A)$ the set of all finite paths, and $\cpaths(\mathcal A)$ the set of
all complete paths of an automaton $\mathcal A$.
\end{dfn}

\noindent Paths are ordered by the prefix relation, which we denote by $\leq$.

Let $\mathcal A$ be a (fully) probabilistic automaton and let $\pi_i$ for $i \geq 0$ be finite paths of $\mathcal A$ all starting in the
same initial state $s_0$ and such that $\pi_i \leq \pi_j$ for $i \leq j$ and $|\pi_i| = i$, for all $i \geq 0$. Then by $\pi = \lim_{i \to
\infty}{\pi_i}$ we denote the infinite complete path with the property that $\pi_i \leq \pi$ for all $i\geq 0$.

\begin{dfn}[Cone]\label{ConeDef}
Let $\mathcal A$ be a (fully) probabilistic automaton and let $\pi \in \fpaths(\mathcal A)$ be given. The cone generated by $\pi$ is the
set of paths
$$C_\pi = \{ \pi'\in \cpaths(\mathcal A) \mid \pi \leq \pi'\}.$$
\end{dfn}

\noindent From now on we fix an initial state. Given a fully probabilistic automaton $\mathcal A$ with an initial state $s_0$, we can
calculate the probability-value denoted by $\psp(\pi)$ of any finite path $\pi$ starting in $s_0$ as follows.
\begin{eqnarray*}
\psp(s_0) & = & 1\\
\psp(\pi \stackrel{a}{\to} s) & = & \psp(\pi)\cdot \mu(a,s) \quad \text{~where~} \last(\pi) \to \mu
\end{eqnarray*}

\noindent Let $\Omega_{\mathcal A} = \cpaths(\mathcal A)$ be the sample space, and let $\mathcal F_{\mathcal A}$ be the smallest
$\sigma$-algebra generated by the cones. The following proposition (see~\cite{Seg95:thesis,Sok05:thesis}) states that $\psp$ induces a
unique probability measure on $\mathcal F_{\mathcal A}$.

\begin{proposition}\label{PMeasProp}
Let $\mathcal A$ be a fully probabilistic automaton and let $\psp$ denote the probability-value on paths. There exists a unique probability
measure on $\mathcal F_{\mathcal A}$ also denoted by $\psp$ such that $\psp(C_\pi) = \psp(\pi)$ for every finite path $\pi$.\qed
\end{proposition}

\noindent This way we are able to measure the probability of certain events described by sets of paths in an automaton with no
nondeterminism. Since our models include nondeterminism, we will first resolve it by means of schedulers or adversaries. Before we define
adversaries note that we can describe the set of all sub-probability distributions on a set $S$ by $\distr(S +1)$. These are functions
whose sum of values on $S$ is not necessarily equal to 1, but it is bounded by 1.

\begin{dfn}[Scheduler]
A scheduler for a probabilistic automaton $\mathcal A$ is a function
$$\xi \colon \fpaths(\mathcal A) \to \distr(A \times \distr(S) + 1)$$ satisfying $\xi(\pi)(a, \mu) \gr 0$ implies $\last(\pi) \stackrel{a}{\to} \mu$,
for each finite path $\pi$.
By $\sched(\mathcal A)$ we denote the set of all schedulers for $\mathcal A$.
\end{dfn}

\noindent Hence, a scheduler according to the previous definition imposes a probability distribution on the possible non-deterministic
transitions in each state. Therefore it is randomized. It is history dependent since it takes into account the path (history) and not only
the current state. It is partial since it gives a sub-probability distribution, i.e., it may halt the execution at any time.

\begin{dfn}[Automaton under scheduler]
A probabilistic automaton $\mathcal A = \langle S, A, \alpha\rangle$ together with a scheduler $\xi$ determine a fully probabilistic
automaton $$\mathcal A_\xi = \langle \fpaths(\mathcal A), A, \alpha_\xi \rangle.$$ Its set of states are the finite paths of $\mathcal A$,
its  initial state is the initial state of $\mathcal A$ (seen as a path with length 1), its actions are the same as those of $\mathcal A$,
and its transition function $\alpha_\xi$ is defined as follows. For any $\pi \in \fpaths(\mathcal A)$,  we have $\alpha_\xi(\pi) \in
\distr(A\times \fpaths(\mathcal A)) + 1$ as
$$\alpha_\xi(\pi)(a,\pi') = \left\{\begin{array}{ll}
        \xi(\pi)(a, \mu)\cdot \mu(s) & \,\,\, \pi' = \pi \stackrel{a,\mu}{\to} s\\
        0 & \,\,\, \text{otherwise}
\end{array}\right.
$$
\end{dfn}

\noindent Given a probabilistic automaton $\mathcal A$ and a scheduler $\xi$, we denote by $\psp_\xi$ the probability measure on sets of
complete paths of the fully probabilistic automaton $\mathcal A_\xi$, as in Proposition~\ref{PMeasProp}. The corresponding $\sigma$-algebra
generated by cones of finite paths of $\mathcal A_\xi$ we denote by $\Omega_\xi$. The elements of $\Omega_\xi$ are measurable sets.

By $\Omega$ we denote the $\sigma$-algebra generated by cones of finite paths of $\mathcal A$ (without fixing the scheduler!) and also call
its elements measurable sets, without having a measure in mind. Actually, we will now show that any scheduler $\xi \in \sched(\mathcal A)$
induces a measure $\psp_{(\xi)}$ on a certain $\sigma$-algebra $\Omega_{(\xi)}$ of paths in $\mathcal A$ such that $\Omega \subseteq
\Omega_{(\xi)}$. Hence, any element of $\Omega$ can be measured by any of these measures $\psp_{(\xi)}$. We proceed with the details.

Define a function $f: \fpaths(\mathcal A_\xi) \to \fpaths(\mathcal A)$ by
\begin{equation}\label{PathsFuncEq}
f(\hat{\pi}) = \last(\hat{\pi})
\end{equation}
for any $\hat{\pi} \in \mathcal A_\xi$. The function $f$ is well-defined since states in $\mathcal A_\xi$ are the finite paths of $\mathcal
A$. Moreover, we have the following property.

\begin{lemma}\label{PrefFLem}
For any $\hat\pi_1, \hat\pi_2 \in \fpaths(\mathcal A_\xi)$ we have
$$\hat\pi_1 \leq \hat\pi_2\quad \iff \quad f(\hat\pi_1) \leq f(\hat\pi_2)$$
where the order on the left is the prefix order in $\fpaths(\mathcal A_\xi)$ and
on the right the prefix order in $\fpaths(\mathcal A)$.
\end{lemma}

\begin{proof}
By the definition of $\mathcal A_\xi$ we have that for $\pi, \pi' \in \fpaths(\mathcal A)$ i.e. states of $\mathcal A_\xi$: $\pi
\stackrel{a}{\leadsto} \pi'$ if and only if $\xi(\pi) \gr 0$ and $\last(\pi) \stackrel{a,\mu}{ s}$ in $\mathcal A$ for some $\mu$ and $s$.
In other words if $\pi \stackrel{a}{\to} \pi'$, then $\pi \leq \pi'$ and $|\pi'| = |\pi| + 1$ i.e. $\pi'$ extends $\pi$ in one step.
Therefore, if we have a path $\pi_0 \stackrel{a_0}{\to} \pi_1 \stackrel{a_1}{\to} \pi_2 \stackrel{a_2}{\to}\cdots$ in $\mathcal A_\xi$ ,
then for all its states: if $i \leq j$, then $\pi_i \leq \pi_j$ and $|\pi_j| = |\pi_i| + (j-i)$. So if $\hat\pi_1, \hat\pi_2 \in
\fpaths(\mathcal A_\xi)$ are such that $\hat\pi_1 \leq \hat\pi_2$, then $\last(\hat\pi_1)$ is a state in $\hat\pi_2$ and therefore we at
once get $\last(\hat\pi_1)\leq \last(\hat\pi_2)$. For the opposite implication, again from the definition we notice that if a path $\hat\pi
\in \fpaths(\mathcal A)$ contains a state $\pi \in \fpaths(\mathcal A)$, then it also contains all prefixes of $\pi$ as states. Hence, if
$\last(\hat\pi_1) \leq \last(\hat\pi_2)$ for $\hat{\pi_1}, \hat{\pi_2} \in \fpaths(\mathcal A_\xi)$, then $\last(\hat\pi_1)$ is a state  in
$\hat\pi_2$ and also all its prefixes are. Since all paths start in the initial state (path), this implies that $\hat\pi_1 \leq \hat\pi_2$.
\qed
\end{proof}

\begin{corollary} \label{InjFCor} The function $f$ defined by~(\ref{PathsFuncEq}) is injective. \qed
\end{corollary}

By Lemma~\ref{PrefFLem} we can extend the function $f$ to
$\hat f: \cpaths(\mathcal A_\xi) \to \cpaths(\mathcal A)$ by
$$\hat f(\hat\pi) = \begin{cases}f(\hat\pi) & \pi \text{~is~finite}\\
               \lim_{i \to infty} f(\hat\pi_i) & \hat\pi_i \leq \pi,\, |\hat\pi_i| = i\end{cases}$$
 The properties from Lemma~\ref{PrefFLem} and Corollary~\ref{InjFCor} continue to hold for the extended function $\hat f$ as well. We will write $f$ for $\hat f$ as well.\\

Recall that $\Omega_\xi$ denotes the $\sigma$-algebra on which the measure $\psp_\xi$ is defined. We now define a family of subsets
$\Omega^{\xi}$ of $\cpaths(\calA)$ by
\begin{equation}\label{OmegaXiEq}
\Omega^\xi = \{ \Pi\in \cpaths(\calA) \mid f^{-1}(\Pi) \in \Omega_\xi.
\end{equation}

The following properties are instances of standard measure-theoretic results.

\begin{lemma}\label{MeasurePropLem}
The family $\Omega^\xi$ is a $\sigma$-algebra on $\cpaths(\calA)$ and by
$$\psp^\xi (\Pi) = \psp_\xi(f^{-1}(\Pi))$$
for $\Pi \in \Omega^\xi$ a measure on $\Omega^\xi$ is defined.
\end{lemma}

Recall that $\Omega$ denotes the $\sigma$-algebra on complete paths of $\calA$ generated by the cones. We show that for any scheduler
$\xi$, $\Omega \subseteq \Omega^\xi$. Hence, the measurable sets (elements of $\Omega$) are indeed measurable by the measure induced by any
scheduler.

\begin{lemma}
For any scheduler $\xi$, $\Omega \subseteq \Omega^\xi$.
\end{lemma}

\begin{proof}
Fix a scheduler $\xi$. Since $\Omega$ is generated by the cones it is enough to show that each cone is in $\Omega^\xi$. Let $C_{\pi_0,
\mathcal A}$ be a cone in $\cpaths(\calA)$ generated by the finite path $\pi_0$, i.e.
$$C_{\pi_0,\mathcal A} = \{\pi \in \cpaths(\calA) \mid \pi_0 \leq \pi\}.$$
We have
$$\hat f^{-1}(C_{\pi_0,\mathcal A}) = \begin{cases}\emptyset & \pi_0 \not\in \hat f(\cpaths(\calA))\\
                    C_{\hat\pi_0, \mathcal A_\xi} & \hat f(\hat\pi_0) = \pi_0\end{cases}.$$
by Lemma~\ref{PrefFLem}. Indeed, let $\pi_0 = \hat f(\hat\pi_0)$. Then
\begin{eqnarray*}
 \hat f^{-1}(C_{\pi_0,\mathcal A}) & = & \{ \hat\pi \in \cpaths(\mathcal A_\xi) \mid \hat f(\hat\pi) \geq \pi_0\}\\
& = & \{ \hat\pi \in \cpaths(\mathcal A_\xi) \mid \hat f(\hat\pi) \geq \hat f(\hat\pi_0)\}\\
(\text{Lem.}~\ref{PrefFLem}) & = & \{ \hat\pi \in \cpaths(\mathcal A_\xi) \mid \hat\pi \geq \hat\pi_0\}\\
& = &  C_{\hat\pi_0, \mathcal A_\xi}
\end{eqnarray*}
\end{proof}

%%%%%%%%%%%%%%%%%%%%%%%%%%%%%%%%%%%%%%%%%%%%%%%%%%%%%%%%%%%%%%%%%%%%%%%%%%%%%%%%%%%%%%%%%%%%%

\noindent We next define two operations on  probabilistic automata used for building composed models out of basic models: parallel
composition and restriction. We compose probabilistic automata in parallel in the style of the process algebra ACP. That is,
asynchronously with communication function given by a semigroup operation on the set of actions. This is the most general way of composing
probabilistic automata in parallel (for an overview see~\cite{SV04:voss}).

\begin{dfn}[Parallel composition]\label{ParCompDef}
We fix an action set $A$ and a communication function $\cdot$ on $A$ which is a partial commutative semigroup. Given two probabilistic
automata $\mathcal A_1 = \langle S_1, A, \alpha_1 \rangle$ and $\mathcal A_2 = \langle S_2, A, \alpha_2 \rangle$ with actions $A$, their
parallel composition is the probabilistic automaton $\mathcal A_1 \parallel \mathcal A_2 = \langle S_1 \times S_2, A, \alpha\rangle$ with
states pairs of states of the original automata denoted by $s_1 \parallel s_2$, the same actions, and transition function defined as
follows. $s_1\parallel s_2 \stackrel{a}{\to} \mu$ if and only if one of the following holds
\begin{itemize}
\item [1.] $s_1 \stackrel{b}{\to} \mu_1$ and $s_2 \stackrel{c}{\to} \mu_2$ for some actions $b$ and $c$ such that $a = b\cdot c$ and\\
\hspace*{3.5cm} $\mu = \mu_1 \cdot \mu_2$ meaning $\mu(t_1 \parallel t_2) = \mu_1(t_1)\cdot \mu_2(t_2)$.
\item [2.] $s_1 \stackrel{a}{\to} \mu'$ where $\mu'(t_1) = \mu(t_1\parallel s_2)$ for all states $t_1$ of the first automaton.
\item [3.] $s_2 \stackrel{a}{\to} \mu'$ where $\mu'(t_2) = \mu(s_1\parallel t_2)$ for all states $t_2$ of the second automaton.
\end{itemize}
Here, 1.\ represents a synchronous joint move of both of the automata, and 2.\ and 3.\ represent the possibilities of an asynchronous move
of each of the automata. In case $s^0_1$ and $s^0_2$ are the initial states of $\mathcal A_1$ and $\mathcal A_2$, respectively, then the
initial state of $\mathcal A_1 \parallel \mathcal A_2$ is $s^0_1 \parallel s^0_2$.
\end{dfn}

\noindent
Often we will use input and output actions like $a?$ and $a!$, respectively, in the style of CCS.%~\cite{}.
In such cases we assume that the communication is defined as hand-shaking $a? \cdot a! = \tau_a$ for $\tau_a$ a special invisible action.\\

\noindent The operation of restriction is needed to prune out some branches of a probabilistic automaton that one need not consider. For
example, we will commonly use restriction to get rid of parts of a probabilistic automaton that still wait on synchronization.

\begin{dfn}[restriction]\label{HidingDef}
Fix a subset $I \subseteq A$ of actions that are in the restricted set. Given an automaton $\mathcal A = \langle S, A, \alpha\rangle$, the
automaton obtained from $\mathcal A$ by restricting the actions in $I$ is $\mathcal R_I(\mathcal A) = \langle S, A \setminus I,
\alpha'\rangle$ where the transitions of $\alpha'$ are defined as follows: $s \stackrel{a}{\to} \mu$ in  $\mathcal R_I(\mathcal A)$ if and
only if $s \stackrel{a}{\to} \mu$ in $\mathcal A$ and $a \not\in I$.
\end{dfn}

%-------------------------------------------------------------------

\noindent We now define bisimilarity - a behaviour equivalence on the states of a probabilistic automaton. For that we first need the
notion of relation lifting.

\begin{dfn}[Relation lifting]
Let $R$ be an equivalence relation on the set of states $S$ of a probabilistic automaton $\calA$. Then $R$ lifts to a relation $\equiv_R$
on the set $\distr(S)$, as follows:
$$\mu \equiv_R \nu \iff \sum_{s \in C} \mu(s) = \sum_{s\in C} \nu(s)$$
for any equivalence class $C \in S/R$.
\end{dfn}

\begin{dfn}[Bisimulation, bisimilarity]
Let $\cal A = \langle S, A, \alpha\rangle$ be a probabilistic automaton. An equivalence $R$ on its set of states $S$ is a bisimulation if
and only if whenever $\langle s, t \rangle \in R$ we have \\

\qquad if $s \stackrel{a}{\to} \mu_s$, then there exists $\mu_t$ such that $t \stackrel{a}{\to} \mu_t$ and $\mu_s \equiv_R \mu_t$.\\

\noindent Two states $s, t \in S$ are bisimilar, notation $s \sim t$ if they are related by some bisimulation relation $R$.
\end{dfn}

\noindent Note that bisimilarity $\sim$ is the largest bisimulation on a given probabilistic automaton $\calA$.
\

\section{Anonymizing Protocols}

\subsection{Dining cryptographers}\label{sec:dining-crypt}

The canonical example of an anonymizing protocol is Chaum's Dining Cryptographers \cite{cha_1988_dining}. In Chaum's introduction to this
protocol, three cryptographers are sitting down to dine in a restaurant, when the waiter informs them that the bill has already been paid
anonymously. They wonder whether one of them has paid the bill in advance, or whether the NSA has done so. Respecting each other's right to
privacy, the carry out the following protocol. Each pair of cryptographers flips a coin, invisible to the remaining cryptographer. Each
cryptographer then reveals whether or not the two coins he say were equal or unequal.  However, if a cryptographer is paying, he states the
opposite. An even number of ``equals'' now indicates that the NSA is paying; an odd number that one of the cryptographers is paying.

\newcommand{\Ftwo}{{\mathbb F}_2}

Formally, Chaum states the result as follows. (Here we are restricting to the case with 3 cryptographers; Chaum's version is more general.)
Here, $\Ftwo$ is the field of two elements.

\begin{thm}
  Let $K$ be a uniformly distributed stochastic variable over
  $\Ftwo^3$. Let $I$ be a stochastic variable over $\Ftwo^3$, taking
  only values in $\{ (1,0,0),\allowbreak (0,1,0),\allowbreak
  (0,0,1),\allowbreak (0,0,0) \}$. Let $A$ be the stochastic variable
  over $\Ftwo^3$ given by $A = (I_1 + K_2 + K_3, K_1 + I_2 + K_3, K_1
  + K_2 + I_3)$. Assume that $K$ and $I$ are independent. Then
  \begin{equation*}
    \forall a \in \Ftwo^3\; \forall i \in \{1,2,3\}:
    \Prob[ I = i ] > 0 \implies \Prob[A = a \vbar I = i] = \tfrac14
  \end{equation*}
  and hence
  \begin{equation*}
    \forall a \in \Ftwo^3\; \forall i \in \{1,2,3\}:
    \Prob[ I = i ] > 0 \implies \Prob[A = a \vbar I = i] = \Prob[A = a].
    \tag*{\qed}
  \end{equation*}
\end{thm}

In terms of the storyline, $K$ represents the coin flips, $I$ represents which cryptographer (if any) is paying, and $A$ represents the
every cryptographer says.  \mnote{I don't know if this version by Chaum should stay. I like its level of abstraction. I also like the way
the we solve our non-deterministic master with a probabilistic choice --- and Chaum exactly models the master as a probabilistic choice of
which you don't know the probabilities. PvR}

\newcommand{\DoPay}{\operatorname{\it pay}}
\newcommand{\Agree}{\operatorname{\it agree}}
\newcommand{\Disagree}{\operatorname{\it disagree}}
\newcommand{\Equal}{\operatorname{\it equal}}
\newcommand{\Unequal}{\operatorname{\it unequal}}

We will now model this protocol as a probabilistic automaton. We will construct it as a parallel composition of seven components: the
Master, who decides who will pay, the three cryptographers Crypt${}_i$, and the three coins Coin$_{i}$. The action $p_i!$ is used by the
Master to indicate to Crypt$_{i}$ that he should pay; the action $n_i!$ to indicate that he shouldn't. If no cryptographer is paying, the
NSA is paying, which is not explicitly modelled here. The coin Coin${}_i$ is shared by Crypt${}_i$ and Crypt${}_{i-1}$ (taking the -1
modulo 3); the action $h_{i,j}!$ represents Coin${}_i$ signalling to Crypt${}_j$ that the coin was heads and similarly $t_{i,j}!$ signals
tails. At the end, the cryptographers state whether or not the two coins they saw were equal or not by means of the actions $a_i!$ (agree)
or $d_i$ (disagree).
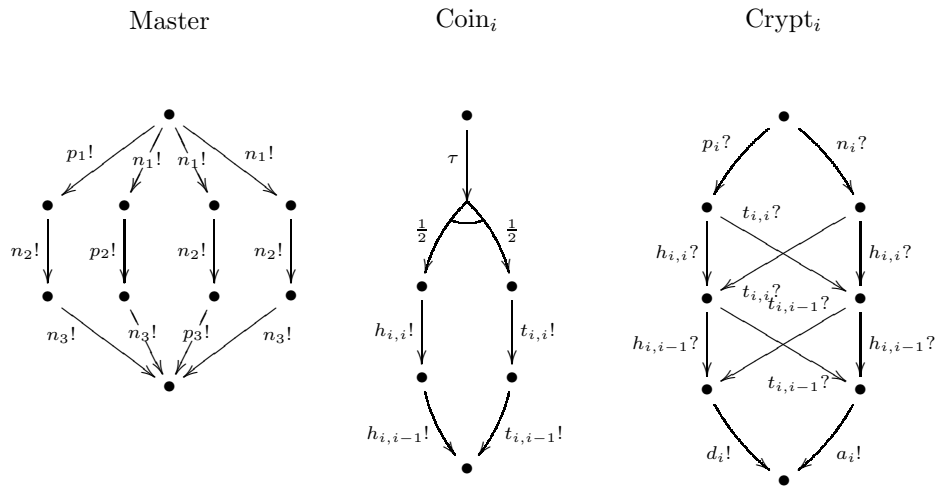
\begin{equation*}
  \xymatrix@C-1.5pc{%
  &&&
  \hbox to 0pt{\hss Master\hss}
  \\
%   &&&
%   \bullet \ar[d]_\tau
%   \\
%   &&&
%   {}
%   \ar[llld]_(0.7){\frac14}|(0.25){}="l"
%   \ar[ld]|(0.5){\frac14}
%   \ar[rd]|(0.5){\frac14}
%   \ar[rrrd]^(0.7){\frac14}|(0.25){}="r"
%   \POS {"l"} \ar@{-}@/_0.1pc/{"r"}
  &&&
  \bullet
  \ar[llld]_(0.6){p_1!}
  \ar[ld]|(0.5){n_1!}
  \ar[rd]|(0.5){n_1!}
  \ar[rrrd]^(0.6){n_1!}
  \\
%   \bullet \ar[d]_{p_1!} 
%   &&
%   \bullet \ar[d]_{n_1!}
%   &&
%   \bullet \ar[d]_{n_1!}
%   &&
%   \bullet \ar[d]_{n_1!}
  \bullet \ar[d]_{n_2!}
  &&
  \bullet \ar[d]_{p_2!}
  &&
  \bullet \ar[d]_{n_2!}
  &&
  \bullet \ar[d]_{n_2!}
  \\
  \bullet \ar[rrrd]_(0.25){n_3!}
  &&
  \bullet \ar[rd]|(0.4){n_3!}
  &&
  \bullet \ar[ld]|(0.4){p_3!}
  &&
  \bullet \ar[llld]^(0.25){n_3!}
  \\
  &&&
  \bullet
}
  \qquad
  \xymatrix@C-1.5pc{%
  &
  \hbox to 0pt{\hss Coin${}_i$\hss}
  \\
  &
  \bullet \ar[d]+0_\tau
  \\
  &
  {}
  \POS[]+0\ar@/_0.25pc/[ld]_{\frac12}|(0.25){}="l"
  \POS[]+0\ar@/^0.25pc/[rd]^{\frac12}|(0.25){}="r"
  \POS {"l"} \ar@{-}@/_0.1pc/{"r"}
  \\
  \bullet \ar[d]_{h_{i,i}!}
  &&
  \bullet \ar[d]^{t_{i,i}!}
  \\
  \bullet \ar@/_0.25pc/[rd]_{h_{i,i-1}!}
  &&
  \bullet \ar@/^0.25pc/[ld]^{t_{i,i-1}!}
  \\
  &
  \bullet
}
  \qquad
  \xymatrix@C-0.5pc{%
  & \hbox to 0pt{\hss Crypt${}_i$\hss}
  \\
  &
  \bullet
  \ar@/_0.25pc/[ld]_{p_i?}
  \ar@/^0.25pc/[rd]^{n_i?}
  \\
  \bullet
  \ar[d]_{h_{i,i}?}
  \ar[rrd]^(0.25){t_{i,i}?}
  &&
  \bullet
  \ar[d]^{h_{i,i}?}
  \ar[lld]^(0.75){t_{i,i}?}
  \\
  \bullet
  \ar[d]_{h_{i,i-1}?}
  \ar[rrd]_(0.75){t_{i,i-1}?}
  &&
  \bullet
  \ar[d]^{h_{i,i-1}?}
  \ar[lld]_(0.25){t_{i,i-1}?}
  \\
  \bullet \ar@/_0.25pc/[rd]_{d_i!}
  &&
  \bullet \ar@/^0.25pc/[ld]^{a_i!}
  \\
  &
  \bullet
}
\end{equation*}
Now DC is the parallel composition of Master, Coin${}_0$, Coin${}_1$, Coin${}_2$, Crypt${}_0$, Crypt${}_1$, and Crypt${}_2$ with all
actions of the form $p_i$, $n_i$, $h_{i,j}$, and $t_{i,j}$ hidden.

Note that in Chaum's version, there is no assumption on the probability distribution of $I$; in our version this is modelled by the fact
that the Master makes a non-deterministic choice between the four options. Since we allow probabilistic schedulers, we later recover all
possible probability distributions about who is paying, just as in the original version. Independence between the choice of the master and
the coin flips ($I$ and $K$ in Chaum's version) comes for free in the automata model: distinct probabilistic choices are always assumed to
be independent.  \mnote{But there actually is a subtle issue if you first have a probabilistic choice and then a non-deterministic choice.
But that's what the whole business with our admissible schedulers is about. PvR}

In Section~\ref{sec:PurelyProbabilisticSystems} we formulate what it means for DC (or more general, for an anonymity automaton) to be
anonymous.

\subsection{Voting}

At a very high level, a voting protocol can be seen as a blackbox that inputs the voters' votes and outputs the result of the vote. For
simplicity, assume the voters vote yes (1) or no (0), do not abstain, and that the numbers of voters is known. The result then is the
number of yes-votes.
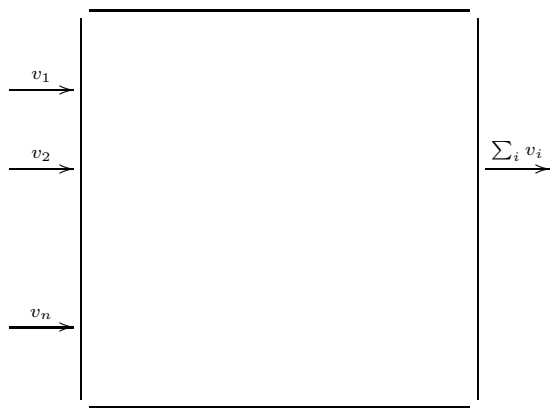
\begin{equation*}
\xymatrix{
  &
  \ar@{-}[ddddd] \ar@{-}[rrrrr]
  &&&&&
  \ar@{-}[ddddd]
  \\
  \ar[r]^{v_1}
  &
  \\
  \ar[r]^{v_2}
  &&&&&&
  \ar[r]^{\sum_i v_i}
  &
  \\
  &
  \\
  \ar[r]^{v_n}
  &
  \\
  &
  \ar@{-}[rrrrr]
  &&&&&
}
\end{equation*}
In such a setting, it is conceivable that an observer has some a-priori knowledge about which voters are more likely to vote yes and which
voters are more likely to vote no. Furthermore, there definitely is a-posteriori knowledge, since the vote result is made public. For
instance, in the degenerate case where all voters vote the same way, everybody's vote is revealed. What we expect here from the voting
protocol is not that the adversary has no knowledge about the votes (since he might already have a-priori knowledge), and also not that the
adversary does not gain any knowledge from observing the protocol (since the vote result is revealed), but rather that observing the
protocol does not augment the adversary's knowledge beyond learning the vote result.

For the purely probabilistic case, this notion of anonymity is formalized in Section~\ref{sec:PurelyProbabilisticSystems}.

\section{Anonymity for Purely Probabilistic Systems}
\label{sec:PurelyProbabilisticSystems}
% We now start by formally defining our notion of anonymity. First, we
% define the notion of anonymity automaton, following (Palamidessi);

% \begin{dfn}
%   Let $\Omega$ be a probability space and let $S \colon \Omega \to T_S$, $O \colon \Omega \to T_O$, and $R \colon \Omega \to T_R$ be stochastic variables. We say that \emph{$S$ is secret with respect to observations $O$ and revelations $R$} if for all $s \in S$, $o \in O$, and $r \in R$,
%   \begin{equation*}
%     \Prob[ S = s \vbar O = o \land R = r] = \Prob[ S = s \vbar R = r].
%   \end{equation*}
% \end{dfn}
% \mnote{Not so clear where this should go. PvR}

This section defines anonymity systems and proposes a definition for anonymity in its simplest configuration, i.e., for purely
probabilistic systems. Purely probabilistic systems are simpler because there is no need for schedulers. Throughout the following sections,
this definitions will be incrementally modified towards a more general setting.
\newcommand{\Otrace}{Otrace}
\begin{dfn}[Anonymity system]
Let $M = \langle S, \Act, \alpha \rangle$ be a fully probabilistic automaton. An \emph{anonymity system} is a triple $\langle M, I, \{ A_i
\}_{i \in I}, \ActO \rangle$ where
\begin{enumerate}
%\item $M$ is a probabilistic automaton,
\item $I$ is the set of user identities, \item $A_i$ is any measurable subset of $\cpaths(M)$ such that $A_i \cap A_j = \emptyset$ for $i
\not= j$. \item $\ActO \subseteq \Act$ is the set of observable actions.
\item $\Otrace(\pi)$ is the sequence of elements in $\ActO$ obtained by removing form $\trace(\pi)$ the elements in $\Act \setminus \ActO$.
\end{enumerate}
Define $O$ as the set of observations, i.e., $O = \{\trace(\pi) \vbar \pi \in \paths(M)\}$. We also define $A = \bigcup_{i \in I} A_i$.
\end{dfn}
Intuitively, the $A_i$s are properties of the executions that the system is meant to hide. For example, in the case of the dining cryptographers $A_i$ would be ``cryptographer $i$ payed''; in a voting scheme ``voter $i$ voted for candidate $c$'', etc. Therefore, for the previous examples, the predicate $A$ would be ``some of the cryptographers payed'' or ``the vote count'' respectively.

Next, we propose a definition of anonymity for a purely probabilistic systems. We deviate from the definition proposed by Bhargava and
Palamidessi~\cite{bp_2005_probabilistic} for what we consider a more intuitive definition: We say that an anonymity system is anonymous if
the probability of seeing a observation is independent of who performed the anonymous action ($A_i$), given that some anonymous action took
place ($A$ happened). The formal definition follows.

\begin{dfn}[Anonymity]
A system $\langle M, I, \{ A_i \}_{i \in I}, \ActO \rangle$ is said to be anonymous if
\begin{align*}
\forall i \in I.\forall o \in O. \Prob[\pi \in A] > 0 \implies & \Prob[\Otrace(\pi) = o \;\land\; \pi \in A_i \vbar \pi \in A] =\\
 & \Prob[\Otrace(\pi) = o \vbar \pi \in A]\, \Prob[\pi \in A_i\vbar \pi \in A].
\end{align*}
\end{dfn}
In the above probabilities, $\pi$ is drawn from the probability space $\Paths(M)$.

The following lemma shows that this definition is equivalent to the one proposed in Bhargava and Palamidessi~\cite{bp_2005_probabilistic}.

\begin{lemma}
A anonymity system is anonymous if and only if
\begin{align*}
\forall i,j \in I.\forall o \in O.(\Prob[\pi \in A_i]>0 \land \Prob[\pi \in A_j]>0) \implies & \Prob[\Otrace(\pi) = o \vbar \pi \in  A_i] =\\
& \Prob[\Otrace(\pi) = o  \vbar \pi \in A_j] \;
\end{align*}
\end{lemma}
\begin{proof}
The only if part is trivial. For the if part we have
\begin{align*}
\Prob[  &\Otrace(\pi) =  o  \vbar \pi \in  A]\, \Prob[\pi \in  A_i \vbar \pi \in  A] \\
&= \Prob[\pi \in  A_i \vbar \pi \in  A] \; \sum_{j \in I} \Prob[\Otrace(\pi) = o  \vbar \pi \in  A_j \cap  A] \; \Prob[\pi \in  A_j  \vbar \pi \in  A]\\
\intertext{\qquad\quad(since $A_i \cap A_j = \emptyset,i \not= j$)}
&= \Prob[\pi \in  A_i \vbar \pi \in  A] \; \sum_{j \in I} \Prob[\Otrace(\pi) = o  \vbar \pi \in  A_j] \; \Prob[\pi \in  A_j  \vbar \pi \in  A]\\
\intertext{\qquad\quad(by definition of $\pi \in  A$)}
&= \Prob[\pi \in  A_i \vbar \pi \in  A] \; \Prob[\Otrace(\pi) = o  \vbar \pi \in  A_i] \;\sum_{j \in I}  \Prob[\pi \in  A_j  \vbar \pi \in  A]\\
\intertext{\qquad\quad(by hypothesis)}
    &= \Prob[\pi \in  A_i \vbar \pi \in  A] \; \Prob[\Otrace(\pi) = o  \vbar \pi \in  A_i] \;\frac{\sum_{j \in I} \Prob[\pi \in  A_j]}{\Prob[\pi \in  A]}\\
\intertext{\qquad\quad(since $A_j \subseteq A$)}
    &= \Prob[\pi \in  A_i \vbar \pi \in  A] \; \Prob[\Otrace(\pi) = o  \vbar \pi \in  A_i] \\
    &= \frac{\Prob[\pi \in  A_i]}{\Prob[\pi \in  A]} \; \frac{\Prob[\Otrace(\pi) = o \land \pi \in  A_i]}{\Prob[\pi \in  A_i]}\\
    &= \Prob[\Otrace(\pi) = o \land \pi \in  A_i  \vbar \pi \in  A]
\intertext{\qquad\quad(since $A_i \subseteq A$)}
\end{align*}
which concludes the proof. \qed
\end{proof}

\section{Anonymity for Probabilistic Systems}

We now try to extend the notion of anonymity to probabilistic automata that are not purely probabilistic, but that still contain some
non-deterministic transitions. \mnote{I think this should simply become the intro to the next section. PvR}

One obvious try is to say that $M$ is anonymous if $M_\xi$ is anonymous for all schedulers $\xi$ of $M$. The following automaton $M$ and
scheduler $\xi$ show that this definition would be problematic.
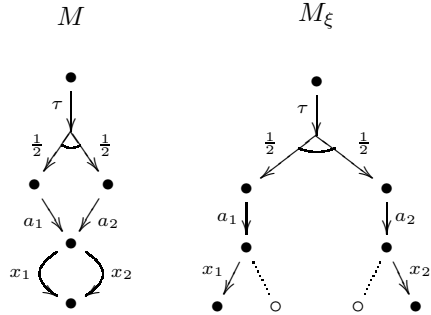
\begin{equation*}
  \xymatrix@C-2pc@R-1pc{
  & M
  \\
  &
  \bullet \ar[d]+0_\tau
  \\
  &
  {}
  \POS[]+0\ar[dl]_{\frac12}|(0.25){}="l"
  \POS[]+0\ar[dr]^{\frac12}|(0.25){}="r"
  \POS {"l"} \ar@{-}@/_0.1pc/{"r"}
  \\
  \bullet
  \ar[dr]_{a_1}
  &&
  \bullet
  \ar[dl]^{a_2}
  \\
  &
  \bullet
  \ar@/_1pc/[d]_{x_1}|(0.25){}="l"
  \ar@/^1pc/[d]^{x_2}|(0.25){}="r"
  \\
  &
  \bullet
}
  \qquad
  \xymatrix@C-2pc@R-1pc{
  &&&
  M_\xi
  \\
  &&&
  \bullet
  \ar[d]+0_\tau
  \\
  &&&
  {}
  \POS[]+0\ar[dll]_{\frac12}|(0.25){}="l"
  \POS[]+0\ar[drr]^{\frac12}|(0.25){}="r"
  \POS {"l"} \ar@{-}@/_0.1pc/{"r"}
  \\
  &
  \bullet
  \ar[d]_{a_1}
  &&&&
  \bullet
  \ar[d]^{a_2}
  \\
  &
  \bullet
  \ar[dl]_{x_1}
  \ar@{.}[dr]
  &&&&
  \bullet
  \ar@{.}[dl]
  \ar[dr]^{x_2}
  \\
  \bullet
  &&
  \circ
  &&
  \circ
  &&
  \bullet
}
\end{equation*}
Here $a_1$ and $a_2$ are invisible actions; they represent which user performed the action that was to remain hidden. The actions $x_1$ and
$x_2$ are observable. Intuitively, because the adversary cannot see the messages $a_1$ and $a_2$, she cannot learn which user actually
performed the hidden action. On the right hand side $M_\xi$ is shown and the branches the scheduler does not take are indicated by dotted
arrows. Now $\Prob_\xi[a_1 \vbar x_1] = 1$, but $\Prob_\xi[a_1] = \frac{1}{2}$, showing that with this particular scheduler $M_\xi$ is not
anonymous.

Note that this phenomenon can easily occur as a consequence of communication non-determinism. For instance, consider the following three
automata and their parallel composition in which $c?$ and $c!$ are hidden. In this example the order of the messages $x_1$ and $x_2$
depends on a race-condition, but a scheduler can make it depend on whether $a_1$ or $a_2$ was taken. I.e., there exists a scheduler $\xi$
such that $\Prob_\xi[x_1 x_2 \vbar a_1] = \Prob_\xi[x_2 x_1 \vbar a_2] = 1$ and hence $\Prob_\xi[x_2 x_1 \vbar a_1] = \Prob_\xi[x_1 x_2
\vbar a_2] = 0$.
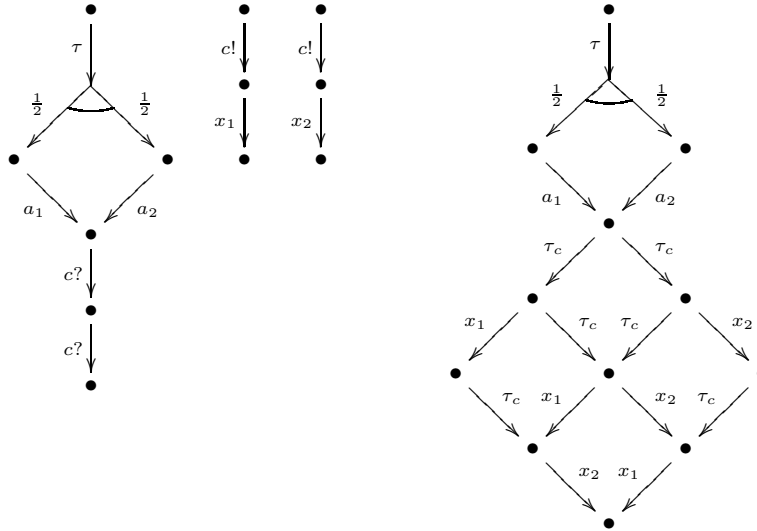
\begin{equation*}
  \xymatrix@C-0.5pc@R-0.5pc{
  &
  \bullet
  \ar[d]+0_\tau
  &&
  \bullet
  \ar[d]_{c!}
  &
  \bullet
  \ar[d]_{c!}
  \\  
  &
  {}
  \POS[]+0\ar[dl]_{\frac12}|(0.3){}="l"
  \POS[]+0\ar[dr]^{\frac12}|(0.3){}="r"
  \POS {"l"} \ar@{-}@/_0.1pc/{"r"}
  &&
  \bullet
  \ar[d]_{x_1}
  &
  \bullet
  \ar[d]_{x_2}
  \\  
  \bullet
  \ar[dr]_{a_1}
  &&
  \bullet
  \ar[dl]^{a_2}
  &
  \bullet
  &
  \bullet
  \\
  &
  \bullet \ar[d]_{c?}
  \\
  &
  \bullet \ar[d]_{c?}
  \\
  &
  \bullet
 }
  \qquad\qquad
  \xymatrix@C-0.5pc@R-0.5pc{
  &&
  \bullet
  \ar[d]+0_\tau
  \\
  &&
  {}
  \POS[]+0\ar[dl]_{\frac12}|(0.3){}="l"
  \POS[]+0\ar[dr]^{\frac12}|(0.3){}="r"
  \POS {"l"} \ar@{-}@/_0.1pc/{"r"}
  \\
  &
  \bullet
  \ar[dr]_{a_1}
  &&
  \bullet
  \ar[dl]^{a_2}
  \\
  &&
  \bullet \ar[dl]_{\tau_c} \ar[dr]^{\tau_c}
  \\
  &
  \bullet \ar[dl]_{x_1} \ar[dr]^{\tau_c}
  &&
  \bullet \ar[dl]_{\tau_c} \ar[dr]^{x_2}
  \\
  \bullet \ar[dr]^{\tau_c}
  &&
  \bullet \ar[dl]_{x_1} \ar[dr]^{x_2}
  &&
  \bullet \ar[dl]_{\tau_c}
  \\
  &
  \bullet \ar[dr]^{x_2}
  &&
  \bullet \ar[dl]_{x_1}
  \\
  &&
  \bullet
  \\
}
\end{equation*}
In fact, the Dining Cryptographers example from Section~\ref{sec:dining-crypt} suffers from exactly the same problem. The order in which
the cryptographers say $\Agree_i$ or $\Disagree_i$ is determined by the scheduler and it is possible to have a scheduler that makes the
paying cryptographer, if any, go last.
\medskip

In \cite{bp_2005_probabilistic}, a system $M$ is called anonymous if for all schedulers $\zeta$, $\xi$, for all observables $o$, and for
all hidden actions $a_i$, $a_j$ such that $\Prob_\zeta[a_i] > 0$ and $\Prob_\xi[a_j] > 0$, $\Prob_\zeta[o \vbar a_i] = \Prob_\xi[o \vbar
a_j]$. This definition, of course, has the same problems as above; in the Dining Cryptographers example in \cite{bp_2005_probabilistic}
this is solved by fixing the order in which the the cryptographers say $\Agree_i$ or $\Disagree_i$. However, also a non-deterministic
choice between two otherwise anonymous systems can become non-anonymous with this definition. For instance, let $P$ be some anonymous
system. For simplicity, assume that $P$ is fully probabilistic (e.g., the Dining Cryptographers with a probabilistic master and a fixed
scheduler) and let $P'$ be a variant of $P$ in which the visible actions have been renamed (e.g., the actions $\Agree_i$ and $\Disagree_i$
are renamed to $\Equal_i$ and $\Unequal_i$). Now consider the probabilistic automaton $M$ which non-deterministically chooses between $P$
and $P'$:
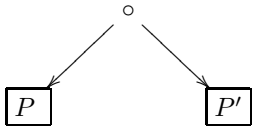
\begin{equation*}\xymatrix{
  & \circ \ar[dl] \ar[dr] \\
  P\phantom{'} \drop\frm{-} && P' \drop\frm{-}
}\end{equation*}
This automaton has only two schedulers: the one that chooses the left branch and then executes $P$ and the one that chooses the right
branch and then executes $P'$. Let's call these schedulers $l$ and $r$ respectively.  Now pick any hidden action $a_i$ and observable $o$
such that $\Prob_l[o \vbar a_i] > 0$. (e.g., $o = \Agree_1 \Disagree_2 \Agree_3$ and $a_i = \DoPay_1$, for which $\Prob[o \vbar \DoPay_1] =
\frac{1}{4})$. Then, nevertheless, $\Prob_r[o \vbar a_i] = 0$, because the observation $o$ cannot occur in $P'$. So, even though
intuitively this system should be anonymous, it is not so according to the definition in \cite{bp_2005_probabilistic}.
\medskip

Every time the problem is that the scheduler has access to information it shouldn't have. When one specifies a protocol by giving a
probabilistic automaton, an implementation of this protocol has to implement a scheduler as well. This is especially obvious if the
non-determinism originates from communication. When we identify schedulers with adversaries, as is common, it becomes clear that the
scheduler should not have access to too much information. In the next section we define a class of schedulers, called
\emph{admissible} schedulers that base their scheduling behavior on the information an adversary actually has access to: the
observable history of the system.

%The term ``admissible'' is chosen because the ``output'' of the scheduler (the scheduling decisions)
%only depend on ``low inputs'' (the observable history) and not on ``high inputs'' (the whole path). \mnote{This whole section is much to
%verbose. PvR}

\section{Admissible Schedulers}

As explained in the previous section, defining anonymity as a condition that should hold true for all possible schedulers is problematic.
It is usual to quantify over all schedulers when showing theoretical properties of systems with both probabilities and non-determinism -
for example we may say ``no matter how the non-determinism is resolved, the probability of an event $X$ is at least $p$''. However, in the
analysis of security protocols, for example with respect to anonymity, we would like to quantify over all possible ``realistic''
adversaries. These are not all possible schedulers as in our theoretic considerations since such a realistic adversary is not able to see
all details of the probabilistic automaton under consideration. Hence, considering that the adversary is any scheduler enables the
adversary to leak information where it normally could not. We call such schedulers interfering schedulers. This way protocols that are
well-known to be anonymous turn out not to be anonymous. One such example is the dining cryptographers protocol explained above. We show
that one gets a better definition of anonymity if one restricts the power of the schedulers, in a realistic way. In this section we define
the type of schedulers with restricted power that we consider good enough for showing anonymity of certain protocols. We call these
schedulers admissible.

Schedulers with restricted power have been treated in the literature. In general, as explained by Segala in~\cite{Seg95:thesis}, a
scheduler with restricted power is given by defining two equivalences, one on the set of finite paths $\equiv_1$ and  another one
$\equiv_2$ on the set of possible transition, in this case $\distr(A\times S)$. Then a scheduler $\xi$ is oblivious relative to
$\langle\equiv_1, \equiv_2\rangle$ if and only if for any two paths $\pi_1, \pi_2$ we have
$$\pi_1 \equiv_1 \pi_2 \Longrightarrow \xi(\pi_1) \equiv_2 \xi(\pi_2).$$

\subsection{Admissible schedulers based on bisimulation}

In this section we specify $\equiv_1$ and $\equiv_2$ and obtain a class of oblivious adversaries that suits the anonymity definition.

Defined $\equiv_1$ on the set of finite paths of an automaton $M$ as,
$$\pi_1 \equiv_1 \pi_2 \iff \big(\trace(\pi_1) = \trace(\pi_2) \land \last(\pi_1) \sim \last(\pi_2)\big).$$
Recall that we defined $\equiv_R$ as the lifting of the equivalence relation $R$ on a set $S$ to an equivalence relation on $\distr(A\times
S)$. For $\equiv_2$ we take the equivalence $\equiv_{\sim}$ on $\distr(A\times S)$. This is well defined since bisimilarity is an
equivalence. Hence, we obtain a class of oblivious schedulers relative $\langle\equiv_1, \equiv_{\sim}\rangle$. These schedulers we call
admissible.

\begin{dfn}[admissible scheduler]
A scheduler is admissible if for any two finite paths $\pi_1$ and $\pi_2$ we have
$$\big(\trace(\pi_1) = \trace(\pi_2) \land \last(\pi_1) \sim \last(\pi_2)\big)\Longrightarrow
\xi(\pi_1) \equiv_{\sim} \xi(\pi_2).$$
\end{dfn}

Intuitively, the definition of a admissible scheduler enforces that in cases when the schedular has observed the same history (given
by the traces of the paths) and is in bisimilar states, it must schedule ``the same'' transitions up to bisimilarity.

%\subsection{Trace} ???

\subsection{Existence}

We now show that admissible schedulers do exist. In fact, we even show that admissible history-independent schedulers exist. A
scheduler $\xi$ is history-independent if it is completely determined by its image of paths of length 0 i.e. if for any path $\pi$ it holds
that $\xi(\pi) = \xi(\last(\pi))$.

\begin{thm}[Existence] There exists a admissible scheduler for
every probabilistic automaton.
\end{thm}

\begin{proof}
Take a probabilistic automaton $M$. We first show that there exists a map $\xi: S \to \distr(A\times S)\cup \{\bot\}$ with the property
that $\xi(s) = \bot$ if and only if $s$ terminates and for all $s, t \in S$, if $s \sim t$, then $\xi(s) \equiv_\sim \xi(t)$.

Consider the set of partial maps
$$\Xi = \left\{\xi: S \hookrightarrow \distr(A\times S)\cup \{\bot\} \left| \begin{array}{ll}
&\xi(s) = \bot \iff s \text{~terminates~},\\
& s \sim t \Longrightarrow \xi(s) \equiv_\sim \xi(t)\\
& \text{~for~} s, t \in dom(\xi)
\end{array}
 \right\}\right..$$
This set is not empty since the unique partial map with empty domain belongs to it.
We define an order $\leq$ on $\Xi$ in a standard way by
$$\xi_1 \leq \xi_2 \iff \big( dom(\xi_1) \subseteq dom(\xi_2) \land \xi_2|_{dom(\xi_1)} = \xi_1 \big).$$

Consider a chain $(\xi_i)_{i \in I}$ in $\Xi$.  Let $\xi = \cup_{i \in I} \xi_i$. This means that $dom(\xi) = \cup_{i\in I} dom(\xi_i)$ and
if $x \in dom(\xi)$, then $\xi(x) = \xi_i(x)$ for $i \in I$ such that $x \in dom(\xi_i)$. Note that $\xi$ is well-defined since $(\xi_i)_{i
\in I}$ is a chain. Moreover, it is obvious that $\xi_i \leq \xi$ for all $i \in I$. We next check that $\xi \in \Xi$.  Let $s, t \in
dom(\xi)$, such that $s \sim t$. Then $s \in dom(\xi_i)$ and $t \in dom(\xi_j)$ for some $i,j \in I$ and either $\xi_1 \leq \xi_2$ or
$\xi_2 \leq \xi_1$. Assume $\xi_1 \leq \xi_2$. Then $s, t \in dom(\xi_j)$ and $\xi_j \in \Xi$ so we have that $\xi_j(s) \equiv_\sim
\xi_j(t)$ showing that $\xi(s) \equiv_\sim \xi(t)$ and we have established that $\xi \in \Xi$.

Hence, every ascending chain in $\Xi$ has an upper bound. By the Lemma of Zorn we conclude that $\Xi$ has a maximal element. Let $\sigma$
be such a maximal element in $\Xi$. We claim that $\sigma$ is a total map. Assume opposite, i.e., there exists $s \in S \setminus
dom(\sigma)$. If there exists a $t \in dom(\sigma)$ such that $s \sim t$ then we define a new partial scheduler $\sigma'$ as follows. If
$\sigma(t) = \bot$ we put $\sigma'(s) = \bot$. If $\sigma(t) = \mu_t$, then, since $t \to \mu_t$ and $s \sim t$, there exists $\mu_s$ such
that $s \to \mu_s$ and $\mu_t \equiv_\sim \mu_s$. In this case we put $\sigma'(s) = \mu_s$. Moreover, put $\sigma'(x) = \sigma(x)$ for $x
\in dom(\sigma)$. Then we have $\sigma'> \sigma$ and $\sigma' \in \Xi$ contradicting the maximality of $\sigma$. Hence $\sigma$ is a total
map.

\sloppypar{Finally, we consider the (history-independent) scheduler $\hat\sigma$ induced by $\sigma$, i.e., defined by $\hat\sigma(\pi) =
\sigma(\last(\pi))$ for any finite path $\pi$. This scheduler is admissible. Namely, given $\pi_1$ and $\pi_2$ such that
$\trace(\pi_1) = \trace(\pi_2)$ and $\last(\pi_1) \sim \last(\pi_2)$ we have, since $\sigma \in \Xi$, that
 $$\hat\sigma(\pi_1) = \sigma(\last(\pi_1)) \equiv_{\sim} \sigma(\last(\pi_2)) = \hat\sigma(\pi_2)$$
which completes the proof}.

\qed
\end{proof}

%
%\begin{dfn}[Action equivalence]
%Let $M$ be a probabilistic automation and let $\mu, \mu' \in \distr(A \times S)$. We say that $\mu$ and $\mu'$ are \emph{action equivalent} if for all $a \in A_V$, $\mu(a) = \mu'(a)$ and
%$\sum_{a \in A_I} \mu(a) = \sum_{a \in A_I}(\mu'(a))$.
%\mnote{$\mu(a) := \sum_{s \in S} \mu(a,s)$, i.e., the probability that the action label is an $a$. PvR}
%\end{dfn}
%
%\begin{thm}
%Let $M$ be a probabilistic automaton. Let $z \colon \fpaths(M) \to \distr(\distr(A \times S))$ such that, for all $\pi \in \fpaths(M)$, $\support(z(\pi)) \subseteq \Delta(\last(\pi))$ and for all action equivalent $\mu, \mu' \in \support(z(\pi))$, $z(\pi)(\mu) = z(\pi)(\mu')$.
%\mnote{Still have to handle $\bot$. PvR}
%Let $\xi$ be the (probabilistic) scheduler on $M$ induced by $z$, i.e.,
%let $\xi \colon \fpaths(M) \to \distr(A \times S)$ be defined by $\xi(\pi) = \sum_{\mu \in \support(z(\pi))} z(\pi)(\mu) \mu$. Then $\xi$ is a nice scheduler on $M$.
%\end{thm}

%\subsection{Obvious Properties}
%
%\subsection{one-step definition}
%
%\section{Anonymity for Mixed Systems}

We are now ready to define anonymity for probabilistic systems, the formal definition follows.
\begin{dfn}[Anonymity]
A system $\langle M, I, \{ A_i \}_{i \in I}, \ActO \rangle$ is said to be anonymous if for all admissible schedulers $\xi$, for all $i \in
I$ and for all $o \in O$
\begin{align*}
\Prob_\xi[\pi \in A] > 0 \implies & \Prob_\xi[\Otrace(\pi) = o \;\land\; \pi \in A_i \vbar \pi \in A] =\\
 & \Prob_\xi[\Otrace(\pi) = o \vbar \pi \in A]\, \Prob_\xi[\pi \in A_i\vbar \pi \in A].
\end{align*}
\end{dfn}

\section{Anonymity Examples}
\label{sec:examples}

In the purely non-deterministic setting, anonymity of a system is often proved (or defined) as follows: take two users $A$ and $B$ and a
trace in which user $A$ is ``the culprit''. Now find a trace that looks the same to the adversary, but in which user $B$ is ``the culprit''
\cite{ho_2003_anonymity,ghrp_2005_anonymity,mvv_2004_anonymity,HasuoK07a}. In fact, this new trace is often most easily obtained by switching the
behavior of $A$ and $B$.

In this section, we make this technique explicit for anonymity in our
setting, with mixed probability and non-determinism.

\begin{dfn}
  Let $M$ be a probabilistic automaton. A map $\alpha \colon S \to S$
  is called an {\em $\ActO$-automorphism} if $\alpha$ induces an automorphism of
  the automation $M_\tau$, which is a copy of $M$ with all actions not in $\ActO$ renamed to $\tau$.
\end{dfn}

The following result generalized the above-mentioned proof technique
that is commonly used for a purely non-deterministic setting.

\begin{thm}
  Consider an anonymity system $(M,I,\ActO)$.  Suppose that for every $i, j \in I$
  there exists a $\ActO$-automorphism $\alpha \colon S \to S$ such that
  $\alpha(A_i) = A_j$. Then the system is anonymous.
\end{thm}

\subsection*{Anonymity of the Dining Cryptographers}

We can now apply the techniques from the previous section to the Dining Cryptographers. Concretely, we show that there exists a
$\ActO$-automorphism exchanging the behaviour of the Crypt${}_1$ and Crypt${}_2$; by symmetry, the same holds for the other two
combinations.

Consider the endomorphisms of Master and Coin${}_2$ indicated in the following figure. The states in the left copy that are not explicitly
mapped (by a dotted arrow) to a state in the right copy are mapped to themselves.
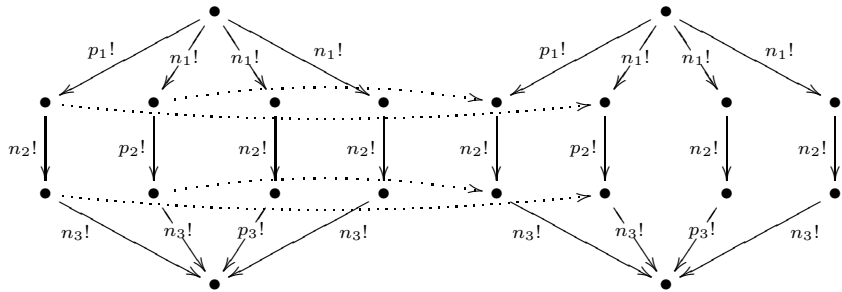
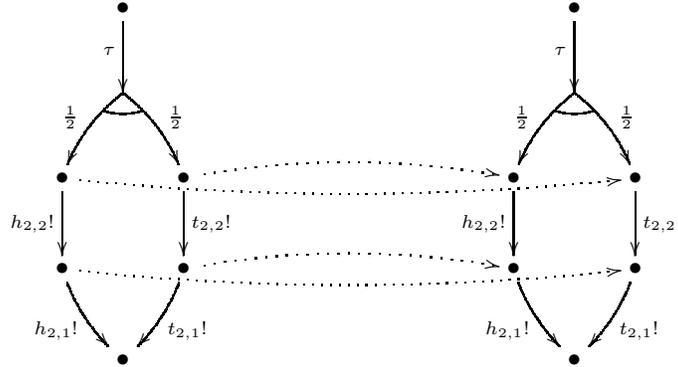
\begin{equation*}
\hspace{3.3cm}\xy
\hspace{-3cm}%
\xymatrix@C=1pc"*"{%
%   &&&
%   \bullet \ar[d]_\tau
%   \\
%   &&&
%   {}
%   \ar[llld]_(0.7){\frac14}|(0.25){}="l"
%   \ar[ld]|(0.5){\frac14}
%   \ar[rd]|(0.5){\frac14}
%   \ar[rrrd]^(0.7){\frac14}|(0.25){}="r"
%   \POS {"l"} \ar@{-}@/_0.1pc/{"r"}
  &&&
  \bullet
  \ar[llld]_(0.6){p_1!}
  \ar[ld]|(0.5){n_1!}
  \ar[rd]|(0.5){n_1!}
  \ar[rrrd]^(0.6){n_1!}
  \\
%   \bullet \ar[d]_{p_1!} 
%   &&
%   \bullet \ar[d]_{n_1!}
%   &&
%   \bullet \ar[d]_{n_1!}
%   &&
%   \bullet \ar[d]_{n_1!}
  \bullet \ar[d]_{n_2!}
  &&
  \bullet \ar[d]_{p_2!}
  &&
  \bullet \ar[d]_{n_2!}
  &&
  \bullet \ar[d]_{n_2!}
  \\
  \bullet \ar[rrrd]_(0.25){n_3!}
  &&
  \bullet \ar[rd]|(0.4){n_3!}
  &&
  \bullet \ar[ld]|(0.4){p_3!}
  &&
  \bullet \ar[llld]^(0.25){n_3!}
  \\
  &&&
  \bullet
}
\POS(60,0)
\xymatrix@C=1pc{%
%   &&&
%   \bullet \ar[d]_\tau
%   \\
%   &&&
%   {}
%   \ar[llld]_(0.7){\frac14}|(0.25){}="l"
%   \ar[ld]|(0.5){\frac14}
%   \ar[rd]|(0.5){\frac14}
%   \ar[rrrd]^(0.7){\frac14}|(0.25){}="r"
%   \POS {"l"} \ar@{-}@/_0.1pc/{"r"}
  &&&
  \bullet
  \ar[llld]_(0.6){p_1!}
  \ar[ld]|(0.5){n_1!}
  \ar[rd]|(0.5){n_1!}
  \ar[rrrd]^(0.6){n_1!}
  \\
%   \bullet \ar[d]_{p_1!} 
%   &&
%   \bullet \ar[d]_{n_1!}
%   &&
%   \bullet \ar[d]_{n_1!}
%   &&
%   \bullet \ar[d]_{n_1!}
  \bullet \ar[d]_{n_2!} \ar@/_0.5pc/@{<.}["*"rr]
  &&
  \bullet \ar[d]_{p_2!} \ar@/^0.5pc/@{<.}["*"ll]
  &&
  \bullet \ar[d]_{n_2!}
  &&
  \bullet \ar[d]_{n_2!}
  \\
  \bullet \ar[rrrd]_(0.25){n_3!} \ar@/_0.5pc/@{<.}["*"rr]
  &&
  \bullet \ar[rd]|(0.4){n_3!} \ar@/^0.5pc/@{<.}["*"ll]
  &&
  \bullet \ar[ld]|(0.4){p_3!}
  &&
  \bullet \ar[llld]^(0.25){n_3!}
  \\
  &&&
  \bullet
}
\endxy
\end{equation*}
\begin{equation*}
\xy
\xymatrix@C=1pc"*"{%
  &
  \bullet \ar[d]+0_\tau
  \\
  &
  {}
  \POS[]+0\ar@/_0.25pc/[ld]_{\frac12}|(0.25){}="l"
  \POS[]+0\ar@/^0.25pc/[rd]^{\frac12}|(0.25){}="r"
  \POS {"l"} \ar@{-}@/_0.1pc/{"r"}
  \\
  \bullet \ar[d]_{h_{2,2}!}
  &&
  \bullet \ar[d]^{t_{2,2}!}
  \\
  \bullet \ar@/_0.25pc/[rd]_{h_{2,1}!}
  &&
  \bullet \ar@/^0.25pc/[ld]^{t_{2,1}!}
  \\
  &
  \bullet
}
\POS(60,0)
\xymatrix@C=1pc{%
  &
  \bullet \ar[d]+0_\tau
  \\
  &
  {}
  \POS[]+0\ar@/_0.25pc/[ld]_{\frac12}|(0.25){}="l"
  \POS[]+0\ar@/^0.25pc/[rd]^{\frac12}|(0.25){}="r"
  \POS {"l"} \ar@{-}@/_0.1pc/{"r"}
  \\
  \bullet \ar[d]_{h_{2,2}!} \ar@/_0.5pc/@{<.}["*"rr]
  &&
  \bullet \ar[d]^{t_{2,2}!} \ar@/^0.5pc/@{<.}["*"ll]
  \\
  \bullet \ar@/_0.25pc/[rd]_{h_{2,1}!} \ar@/_0.5pc/@{<.}["*"rr]
  &&
  \bullet \ar@/^0.25pc/[ld]^{t_{2,1}!} \ar@/^0.5pc/@{<.}["*"ll]
  \\
  &
  \bullet
}
\endxy

\end{equation*}
Also consider the identity endomorphism on Crypt${}_i$ (for $i = 0, 1, 2$) and on Coin${}_i$ (for $i = 0, 1$). Taking the product of these
seven endomorphisms, we obtain an endomorphism $\alpha$ of DC.

\mnote{We probably also need to map Crypt${}_1$ and Crypt${}_2$; now $\alpha$ has nice properties that make DC anonymous. PvR}

%\section{Conclusions}

%\bibliographystyle{alpha}
%\bibliography{source,prob.anon.rev,abbrev-string}

\begin{thebibliography}{GHvRP05}

\bibitem[Bel57]{Bellman_1957_markov}
R.~Bellman.
\newblock A markovian decision process.
\newblock {\em Journal of Mathematics and Mechanics}, 6, 1957.

\bibitem[BP05]{bp_2005_probabilistic}
Mohit Bhargava and Catuscia Palamidessi.
\newblock Probabilistic anonymity.
\newblock In Mart{\'\i}n Abadi and Luca de~Alfaro, editors, {\em Concurency
  Theory, 16th International Conference (CONCUR '05)}, volume 3653 of {\em
  Lecture Notes in Computer Science}, pages 171--185. Springer, 2005.

\bibitem[Cha88]{cha_1988_dining}
David Chaum.
\newblock The dining cryptographers problem: Unconditional sender and recipient
  untraceability.
\newblock 1(1):65--75, 1988.

\bibitem[Che06]{Che06:thesis}
Ling Cheung.
\newblock {\em Reconciling nondeterministic and probabilistic choices}.
\newblock PhD thesis, RU Nijmegen, 2006.

\bibitem[GHvRP05]{ghrp_2005_anonymity}
Flavio~D. Garcia, Ichiro Hasuo, Peter van Rossum, and Wolter Pieters.
\newblock Provable anonymity.
\newblock In Ralf K{\"u}sters and John Mitchell, editors, {\em Proceedings of
  the 2005 {ACM} Workshop on Formal Methods in Security Engineering (FMSE
  '05)}, pages 63--72. ACM, 2005.

\bibitem[HK07]{HasuoK07a}
Ichiro Hasuo and Yoshinobu Kawabe.
\newblock Probabilistic anonymity via coalgebraic simulations.
\newblock In {\em European Symposium on Programming (ESOP '07)}, volume 4421 of
  {\em Lecture Notes in Computer Science}, pages 379--394. Springer, Berlin,
  2007.

\bibitem[HO03]{ho_2003_anonymity}
Joseph Halpern and Kevin O'Neill.
\newblock Anonymity and information hiding in multiagent systems.
\newblock In {\em 16th {IEEE} Computer Security Foundations Workshop (CSFW
  '03)}, pages 75--88, 2003.

\bibitem[MVdV04]{mvv_2004_anonymity}
S.~Mauw, J.~Verschuren, and E.P. de~Vink.
\newblock A formalization of anonymity and onion routing.
\newblock In P.~Samarati, P.~Ryan, D.~Gollmann, and R.~Molva, editors, {\em
  Proceedings of Esorics 2004}, volume 3193 of {\em Lecture Notes in Computer
  Science}, pages 109--124, 2004.

\bibitem[Seg95]{Seg95:thesis}
R.~Segala.
\newblock {\em Modeling and verification of randomized distributed real-time
  systems}.
\newblock PhD thesis, MIT, 1995.

\bibitem[SL94]{SL94:concur}
R.~Segala and N.A. Lynch.
\newblock Probabilistic simulations for probabilistic processes.
\newblock In {\em Proc.\ Concur'94}, pages 481--496. LNCS~836, 1994.

\bibitem[Sok05]{Sok05:thesis}
A.~Sokolova.
\newblock {\em Coalgebraic analysis of probabilistic systems}.
\newblock PhD thesis, TU Eindhoven, 2005.

\bibitem[SV04]{SV04:voss}
A.~Sokolova and E.P.~de Vink.
\newblock Probabilistic automata: system types, parallel composition and
  comparison.
\newblock In C.~Baier, B.R. Haverkort, H.~Hermanns, J.-P. Katoen, and
  M.~Siegle, editors, {\em Validation of Stochastic Systems: A Guide to Current
  Research}, pages 1--43. LNCS 2925, 2004.

\end{thebibliography}

\end{document}